\documentclass[12pt]{article}
\usepackage{titlesec}

\titleformat{\section}{\normalfont\Large\bfseries}{\thesection}{0.3em}{}
\titleformat{\subsection}{\normalfont\large\bfseries}{\thesubsection}{0.3em}{}
\titleformat{\subsubsection}{\normalfont\normalsize\bfseries}{\thesubsubsection}{0.3em}{}
\usepackage[nosort,noadjust]{cite}
\usepackage{amsmath,amsthm,amsfonts,amssymb,amscd,mathrsfs,slashed,graphicx,braket}
\usepackage{hyperref}
\usepackage[dvipsnames]{xcolor}
\usepackage{bm}
\hypersetup{
	colorlinks=true,
	citecolor=blue,
	linkcolor=blue,
	urlcolor=blue
}

\newlength{\xtrawidth}
\newlength{\xtraheight}
\numberwithin{equation}{section}
\numberwithin{table}{section}
\numberwithin{figure}{section}
\def\norm#1{\left|\!\left| {#1} \right|\!\right|}

\begin{document}
\pagenumbering{Alph}
\begin{titlepage}
\vspace*{-2.5cm}
\begin{center}
\hfill MITP/24-046
\vskip 0.6in
{\LARGE\bf{Minimally Extended Current Algebras}}\\[3ex]
{\LARGE\bf{of Toroidal Conformal Field Theories}}
\vskip 0.5in
{\bf
Hans Jockers$^{1,a}$,
Maik Sarve$^{1,b}$,
Ida G.~Zadeh$^{2,c}$}
\vskip 0.3in
{\it $^{1}\,$PRISMA+ Cluster of Excellence \& Mainz Institute for Theoretical Physics\\
Johannes Guttenberg-Universit\"at\\
Staudinger Weg 7, 55128 Mainz, Germany}
\vskip 0.2in
{\it $^{2}\,$Mathematical Sciences and STAG Research Centre\\
University of Southampton, Highfield\\
Southampton SO17 1BJ, UK}
\vskip 0.2in
${}^a$\,{\tt jockers@uni-mainz.de}\\
${}^b$\,{\tt msarve@uni-mainz.de}\\
${}^c$\,{\tt ida.ghazvini-zadeh@soton.ac.uk}
\end{center}
\vskip 0.1in
\begin{center} {\bf Abstract} \end{center}

It is well-known that families of two-dimensional toroidal conformal field theories possess a dense subset of rational toroidal conformal field theories, which makes such families an interesting testing ground about rationality of conformal field theories in families in general. Rational toroidal conformal field theories possess an extended chiral and anti-chiral algebra known as W-algebras. Their partition functions decompose into a finite sum of products of holomorphic and anti-holomorphic characters of these W-algebras. Instead of considering these characters, we decompose the partition functions into products of characters of minimal extensions of $\widehat{\mathfrak{u}}(1)$ current algebras, which already appear for rational conformal field theories with target space $S^1$. We present an explicit construction that determines such decompositions. While these decompositions are not unique, they are universal in the sense that any rational toroidal conformal field theory with a target space torus of arbitrary dimension admits such decompositions. We illustrate these decompositions with a few representative examples of rational toroidal conformal field theories with two- and three-dimensional target space tori. 

\vfill
\noindent April 2024
\end{titlepage}
\pagenumbering{gobble}
\tableofcontents

\newpage
\pagenumbering{arabic}

\section{Introduction and Summary} \label{sec:Intro}
The primary fields of two-dimensional rational conformal field theories (RCFTs) assemble themselves into finitely many blocks of a symmetry algebra. For two-dimensional minimal model conformal field theories (CFTs) this symmetry algebra is the Virasoro algebra and these minimal models consist of finitely many primary fields. More general RCFTs possess infinitely many primary fields that assemble themselves into a finite number of blocks with respect to the extended chiral algebra of the CFT.\footnote{For a review on RCFTs and their properties, see for instance refs.~\cite{DiFrancesco:1997nk,Moore:1989vd,Ginsparg:1988ui,Blumenhagen:2009zz}.}

For a given continuous family of two-dimensional CFTs each member corresponds to a point in the moduli space $\cal M$ of that family of theories. A natural question to ask is which members of this family furnish RCFTs or more generally what is the set $\mathcal{R}$ of RCFTs and its properties in the moduli space $\mathcal{M}$? An important property for a given family of two-dimensional conformal fields theories is whether the set $\mathcal{R}$ is dense in $\mathcal{M}$ \cite{Moore:1998pn,Wendland:2000ye,Hosono:2002yb,Gukov:2002nw,Chen:2005gm,Benjamin:2020flm,Kidambi:2022wvh,Okada:2022jnq}, because the denseness of the set $\mathcal{R}$ would imply that any CFT in $\mathcal{M}$ can be approximated arbitrarily well in terms of a RCFT from the set $\mathcal{R}$. 

Determining the set $\mathcal{R}$ of RCFTs for a given family $\mathcal{M}$ of two-dimensional CFTs is in general a difficult question to answer \cite{Moore:1998pn,Gukov:2002nw,Benjamin:2020flm,Kidambi:2022wvh,Okada:2022jnq}. For families $\mathcal{M}$ of toroidal CFTs, which are free two-dimensional CFTs of scalar fields that parametrize a toroidal target space $T^N$, rationality can be checked explicitly and therefore the set $\mathcal{R}$ of RCFTs is known \cite{Anderson:1988to,Wendland:2000ye,Hosono:2002yb}. Furthermore, the set $\mathcal{R}$ is dense in the moduli space $\mathcal{M}$ of toroidal CFTs \cite{Moore:1998pn,Wendland:2000ye,Hosono:2002yb}.

For families $\mathcal{M}$ of two-dimensional CFTs that are more general than toroidal CFTs --- such as families of infrared fixed-points of two-dimensional non-linear sigma models with Ricci-flat target spaces and supersymmetric extensions thereof --- not much is known about the set $\mathcal{R}$ of RCFTs \cite{Gukov:2002nw,Benjamin:2020flm,Kidambi:2022wvh,Okada:2022jnq}. For instance, the question about the denseness of RCFTs in the moduli spaces $\mathcal{M}$ is not answered in general. Nevertheless, some of the two-dimensional CFTs have submoduli spaces that are described by orbifolds or orientifolds of two-dimensional toroidal CFTs \cite{Harvey:1987da,Sen:1997gv,Wendland:2000ye,Wendland:2000ry,Wendland:2001ks,Nahm:1999ps,Lust:2006zh,Reffert:2006du}. Along such subfamilies the rationality is readily determined from the unorbifolded family of toroidal CFTs. From such subfamiles one can hope to infer properties of RCFTs even in the large family $\mathcal{M}$ of CFTs, for instance by applying methods of conformal perturbation theory \cite{Keller:2019yrr,Benjamin:2020flm,Keller:2023ssv}. 

Prominent examples of families of two-dimensional CFTs with subfamilies of orbifolds of toroidal CFTs arise from (supersymmetric extensions of) two-dimensional non-linear sigma models with polarized K3~surfaces or higher dimensional Calabi--Yau manifolds as their target spaces \cite{Wendland:2000ry,Wendland:2001ks,Lust:2006zh,Reffert:2006du}. Due to the even-dimensionality of Calabi--Yau manifolds, the orbifold toroidal subfamilies also arise from even-dimensional target space tori. Less studied examples are families of non-linear sigma models with seven-dimensional $G_2$-target spaces~\cite{Shatashvili:1994zw,Roiban:2001cp}.  As moduli spaces of $G_2$-manifolds are less understood than moduli spaces of Calabi--Yau manifolds, families of two-dimensional CFTs based on non-linear sigma models with $G_2$-target space manifolds are even more challenging to explore. Nevertheless, there are also examples of families of $G_2$-manifolds with orbifolds of seven-dimensional tori~$T^7$ as submoduli spaces \cite{MR1424428,MR1787733}. In  the associated family $\mathcal{M}$ of two-dimensional CFTs along the sublocus of orbifolds of $T^7$, rationality can again be deduced from the rationality of toroidal CFTs with seven-dimensional toroidal target spaces.

The aim of this note is to describe rational toroidal CFTs explicitly by decomposing their partition functions into finite sums of products of characters of certain minimal extensions of the $\widehat{\mathfrak{u}}(1)$~current algebras. For rational toroidal CFTs with target space tori $T^D$, $D>1$, such a decomposition is finer than the decomposition into characters with respect to their whole extended chiral algebra, but it is not unique. As explained in the main text, the decomposition depends on a choice of sublattice of the even self-dual lattice $\Gamma_{D,D}$ of signature $(D,D)$ of the CFT. While the decomposition of rational toroidal partition functions into characters of their whole extended chiral algebra is well-studied (in particular in the context of even-dimensional target space tori), the systematic decomposition into characters of minimal extensions of $\widehat{\mathfrak{u}}(1)$~current algebras is less explored in the literature.\footnote{See ref.~\cite{Furuta:2023xwl}, where similar decompositions are considered in a different context.} Our motivation for considering such decompositions into blocks of extended $\widehat{\mathfrak{u}}(1)$~current algebras is two-fold. On the one hand, such minimal decompositions are universally applicable for target space tori of any dimension, as long as the target space torus is related to a rational toroidal CFT. On the other hand, as these decompositions do not rely on the symmetry of the entire extended chiral algebra, such finer decompositions can be useful to explicitly describe specific orbifolds or orientifold of such rational toroidal CFTs. We hope that the technical result of this short note proves useful in testing some of the deeper questions about RCFTs.

The characters of the extension of the $\widehat{\mathfrak{u}}(1)$~current algebras relevant for this work already appear for the two-dimensional CFTs of a real free boson~$\phi$ on a circle and are discussed in detail, for instance in refs.~\cite{DiFrancesco:1997nk,Ginsparg:1988ui,Blumenhagen:2009zz}. The partition function for this CFT reads\footnote{In this note we use the convention $\alpha'=2$.}
\begin{equation} \label{eq:ZS1}
 Z_{S^1}(\tau; R) = \frac{1}{|\eta(\tau)|^2} \sum_{m,n \in \mathbb{Z}} 
   q^{\frac{1}{2}(\frac{m}{R}+\frac{R n}{2})^2} \bar{q}^{\frac{1}{2}(\frac{m}{R}-\frac{R n}{2})^2} \ , \qquad q = e^{2\pi i \tau} \ ,
\end{equation}
in terms of the modular parameter $\tau$ in the Siegel upper half plane, the Dedekind eta function $\eta(\tau)$, and the radius $R$ of the circle $S^1$. It is well known that for the radius $R = \sqrt{\frac{2p'}{p}}$ with positive co-prime integers $p$ and $p'$ the CFT is rational, because the $\widehat{\mathfrak{u}}(1)$ current algebra of the free boson generated by the chiral primary field $\partial\phi$ is extended by the primary fields $e^{\pm i \sqrt{2p p'} \phi}$, see for instance refs.~\cite{DiFrancesco:1997nk,Ginsparg:1988ui}. Following ref.~\cite{DiFrancesco:1997nk}, we denote this extension of the $\widehat{\mathfrak{u}}(1)$~current algebra by $\widehat{\mathfrak{u}}(1)_{pp'}$, and we often refer to them as minimal extensions of the $\widehat{\mathfrak{u}}(1)$~current algebra. Then the partition function~\eqref{eq:ZS1} simplifies to \cite{DiFrancesco:1997nk}
\begin{equation} \label{eq:ZS1ratDec}
    Z_{S^1}^\text{rat}(\tau; p,p') = \sum_{\lambda \in \{0,1, \dots, 2pp^{\prime}-1\}} 
    \mathcal{K}_{\omega \lambda,pp'}(\tau) \;\overline{\mathcal{K}_{\lambda,pp'}(\tau)} \ ,
\end{equation}
where $\omega = p r + p' s$ for a B\'ezout pair $(r,s)$ obeying $pr -p' s=1$. Furthermore, $\mathcal{K}_{\lambda,\alpha}(\tau)$ are the irreducible characters of $\widehat{\mathfrak{u}}(1)_{\alpha}$  defined for any positive integer $\alpha$ as
\begin{equation} \label{eq:DefK}
  \mathcal{K}_{\lambda,\alpha}(\tau) = \frac{1}{\eta(\tau)} \sum_{n\in \mathbb{Z}} q^{\frac{(2\alpha n+\lambda)^2}{4\alpha}} \ ,  \qquad \lambda \in
  \mathbb{Z} \ ,
\end{equation}
with the symmetry properties
\begin{equation} \label{eq:Ksym}
  \mathcal{K}_{\lambda,\alpha}(\tau) = \mathcal{K}_{\lambda +2\alpha ,\alpha}(\tau)  \ , \qquad
  \mathcal{K}_{\lambda,\alpha}(\tau) = \mathcal{K}_{-\lambda,\alpha}(\tau) \ .
\end{equation}

The main result of this note concerns the decomposition of the partition function of rational toroidal CFTs with a target space $T^D$ into the characters~\eqref{eq:DefK}
\begin{equation} \label{eq:ZTNrat}
   Z_{T^D}^\text{rat}(\tau) = \sum_{\lambda \in I} \mathcal{K}_{\lambda_1, \alpha_1}(\tau)\ldots \mathcal{K}_{\lambda_D, \alpha_D}(\tau)
   \;\overline{\mathcal{K}_{\tilde\lambda_1, \tilde\alpha_1}(\tau)}\ldots \overline{\mathcal{K}_{\tilde\lambda_D, \tilde\alpha_D}(\tau)} \ ,
\end{equation}
for suitable positive integers $\alpha_1,\ldots,\alpha_D,\tilde\alpha_1,\ldots,\tilde\alpha_D$ and for a finite set $I$ of $2D$-tuples $(\lambda_1,\ldots,\lambda_D, \tilde\lambda_1,\ldots,\tilde\lambda_D)$. 

This decomposition~\eqref{eq:ZTNrat} is based on the existence of an orthogonal (not necessarily primitive) sublattice $O_L \oplus O_R$ of the even self-dual charge lattice $\Gamma_{D,D}$ of signature $(D,D)$ of the rational toroidal CFT with target space $T^D$. We show that the summands in the decomposition~\eqref{eq:ZTNrat} are in one-to-one corresponds with the cosets of $\Gamma_{D,D}/O_L \oplus O_R$. As a consequence there is always a universal summand in each expansion~\eqref{eq:ZTNrat} that is attributed to the sublattice $O_L \oplus O_R$ itself. We present a construction that explicitly determines such expansions for any dimension $D$. As explained in the main text our construction amounts to identifying a $2D$-tuple of positive integers $(\alpha_1, \ldots, \alpha_D, \tilde \alpha_1, \ldots, \tilde\alpha_D)$ and a $2D\times 2D$-dimensional integral matrix $\bm{H}$ from the partition function of the rational toroidal CFT with target space $T^D$. After determining the Hermite normal form of the matrix $\bm{H}$ --- for which well-developed algorithms are implemented in most computer algebra packages --- we can unambiguously read off the expansion~\eqref{eq:ZTNrat}. We illustrate our findings with explicit examples, which for ease of presentation we limit to rational toroidal CFTs with two- and three-dimensional toroidal target spaces.

The structure of this work is as follows: In Section~\ref{sec:TDCFTs} we introduce the partition functions for toroidal CFTs with a $D$-dimensional target space $T^D$, and we review the known criteria for such theories to be rational. We show that the partition function of any rational toroidal CFT can be decomposed into the form~\eqref{eq:ZTNrat}, and we compare this decomposition to the decomposition into characters of the whole extended chiral algebra of rational toroidal CFTs. In Section~\ref{sec:T2Par} we present our construction to explicitly calculate the decomposition~\eqref{eq:ZTNrat}. For ease of presentation we first focus on rational toroidal CFTs with target space $T^2$. We explain that the derived construction for two-dimensional toroidal target spaces carries over --- without any further modification --- to rational toroidal CFTs with target space $T^D$ for any dimension $D$. In Section~\ref{sec:Examples} we present explicit examples of decompositions of rational toroidal partition functions into the form~\eqref{eq:ZTNrat} for two-dimensional and three-dimensional toroidal target spaces.

\section{Rational Toroidal Conformal Field Theories} \label{sec:TDCFTs}
In this section we first review the partition functions of CFTs with toroidal target spaces $T^D$ and the well-known conditions for them to be rational. In subsection \ref{sec:RTCFTpar} we show that the rationality property allows us to rewrite the partition functions of these theories in a form which admits an expansion into characters of $\widehat{\mathfrak{u}}(1)_\alpha$ introduced in the introduction --- see eq. \eqref{eq:DefK}. In subsection \ref{sec:RTCFTcomp} we compare the expansion into such characters with the more conventional expansion into characters of the whole extended algebra of rational toroidal CFTs.

\subsection{Rational toroidal conformal field theory partition functions} \label{sec:RTCFTpar}
To set the stage let us introduce the partition function of toroidal CFT. Let us realize the target space torus $T^D$ of such toroidal CFTs in terms of the quotient
\begin{equation}
  T^D  \simeq \mathbb{R}^D / \Lambda \ ,
\end{equation}  
where $\Lambda$ is a $D$-dimensional lattice in $\mathbb{R}^D$, which induces its pairing from the standard scalar product of $\mathbb{R}^D$. The lattice $\Lambda$ is referred to as the torus lattice, and we explicitly represent it in terms of a $D\times D$-matrix $\bm{\Lambda}$, whose columns are the lattice generators.  The partition function of toroidal CFT with target space~$T^D$ takes the well-known general form \cite{Polchinski:1998rq,Wendland:2000ye,Blumenhagen:2009zz,Blumenhagen:2013fgp}
\begin{equation} \label{eq:toruspart}
	Z_{T^{D}}(\tau) = \frac{1}{|\eta(\tau)|^{2D}} \sum_{\bm{p}=\bm{p_L}+\bm{p_R} \in \Gamma_{D,D}} q^{\frac{1}{2}\bm{p_L}^2} \bar{q}^{\frac{1}{2}\bm{p_R}^2} \ ,
	\qquad q=e^{2\pi i \tau} \ .
\end{equation}
Here $\Gamma_{D,D}$ is the even self-dual lattice of signature $(D,D)$ for the momenta $\bm p \in \Gamma_{D,D}$, which decompose into left- and right-moving momenta as $\bm{p}=\bm{p_L} +\bm{p_R}$. The toroidal CFT has an infinite number of Virasoro primary fields, because each point in the lattice $\Gamma_{D,D}$ gives rise to a Virasoro primary of conformal weight $(h_L,h_R) = (\frac{\bm{p_L}^2}2,\frac{\bm{p_R}^2}2)$. In addition to the target space $T^D$ described in terms of the lattice $\Lambda$ --- encoding the metric $\bm{G}$ of the torus $T^D$ as $\bm{G} = \bm{\Lambda}^T \bm{\Lambda}$ --- the toroidal CFT is entirely determined by a choice of anti-symmetric $B$-field $\bm{\widetilde{B}}$. Following ref.~\cite{Wendland:2000ye}, this data parametrizes the left-moving and right-moving momenta as
\begin{equation} \label{plprtorus}
\bm{p_L} = \frac{1}{\sqrt{2}}  \left(\bm{\mu} - \bm{\widetilde{B}} \bm{\lambda} + \bm{\lambda}  \right) \ , \qquad
\bm{p_R} =\frac{1}{\sqrt{2}} \left(\bm{\mu} - \bm{\widetilde{B}} \bm{\lambda} - \bm{\lambda}  \right) \ ,
\end{equation}
with $(\bm{\mu}, \bm{\lambda}) \in \Lambda^{*}  \oplus \Lambda$ in terms of the $D$-dimensional lattice $\Lambda^{*}$ dual to the $D$-dimensional torus lattice $\Lambda$.

The toroidal CFT with target space $T^D$ and anti-symmetric tensor field $\bm{\widetilde{B}}$ is rational if and only if all conformal weights $(h_L, h_R)$ of the primary fields are rational \cite{Anderson:1988to, Moore:1989vd,Wendland:2000ye,Gukov:2002nw}. This condition is equivalent to the requirement that the metric and the anti-symmetric $B$-field represented in terms of the matrices $\bm{G} = \bm{\Lambda}^T \bm{\Lambda}$ and $\bm{B}= \bm{\Lambda}^T \bm{\widetilde{B}} \bm{\Lambda}$ are rational as well, namely
\begin{equation} \label{eq:GBrat}
  \bm{G} \in \operatorname{Sym}_D(\mathbb{Q})\ , \qquad
  \bm{B} \in \operatorname{Skew}_D(\mathbb{Q}) \ ,
\end{equation}
where $\operatorname{Sym}_D(\mathbb{Q})$ and $\operatorname{Skew}_D(\mathbb{Q})$ denote the spaces of symmetric and skew-symmetric $D\times D$~matrices with values in the rational numbers $\mathbb{Q}$. As a result any partition function~\eqref{eq:toruspart} of a rational toroidal CFT has the general form
\begin{equation}\label{eq:toruspart2}
	Z_{T^D}^\text{rat}(\tau ) = \frac{1}{|\eta(\tau)|^{2D}} \sum_{\bm{m}, \bm{n}  \in \mathbb{Z}^D} 
	\prod_{i=1}^D  q^{\frac{1}{4a_i} (\bm{a}_i^T \bm{m} + \bm{b}_i^T \bm{n})^2} 
	\prod_{j=1}^D \bar{q}^{\frac{1}{4\tilde a_j} (\bm{\tilde a}_i^T \bm{m} + \bm{\tilde b}_i^T \bm{n})^2}  \ ,
\end{equation}
with integers $a_i, \tilde a_j \in  \mathbb{Z}_{>0}$ and integral vectors $\bm{a}_i, \bm{\tilde a}_j, \bm{b}_i, \bm{\tilde b}_j \in  \mathbb{Z}^D$ for $i,j=1,\ldots,D$. 

Another consequence of the rationality of the toroidal CFT is that the there is a sublattice $\Gamma_{L,0} \oplus \Gamma_{0,R}$ of the even self-dual $\Gamma_{D,D}$, where the summands $\Gamma_{L,0}$ and $\Gamma_{0,R}$ are $D$-dimensional even mutually orthogonal lattices of left- and right-moving momenta $\bm{p_L}$ and $\bm{p_R}$, see for instances refs.~\cite{Wendland:2000ye,Gukov:2002nw}. Moreover, the lattices $\Gamma_{L,0}$ and $\Gamma_{0,R}$ have sublattice $O_L$ and $O_R$ of finite index respectively, which both have mutually orthogonal generators. We denote the generators of $O_L$ and $O_R$ by $\bm{o}_1,\ldots,\bm{o}_D$ and $\bm{\tilde o}_1,\ldots,\bm{\tilde o}_D$, respectively. They all have an even norm (length squared) due to the evenness of $\Gamma_{D,D}$. 
We then arrive at the following hierarchy of finite index sublattices
\begin{equation} \label{eq:OrthSub}
  O_L \oplus O_R \subset \Gamma_{L,0} \oplus \Gamma_{0,R} \subset \Gamma_{D,D} \ .
\end{equation}
 
 Restricting the summation in the partition function~\eqref{eq:toruspart2} to the sublattice $O_L \oplus O_R$ yields the contribution
\begin{equation} \label{eq:Zsublattice}
  Z^{O_L \oplus O_R}_{T^D}(\tau) = \frac1{|\eta(\tau)|^{2D}} \prod_{i=1}^D \left( \sum_{k\in\mathbb{Z}} q^{\alpha_i k^2} \right)
  \left( \sum_{k\in\mathbb{Z}} {\bar q}^{\tilde \alpha_i k^2} \right) 
  = \prod_{i=1}^D \mathcal{K}_{0,\alpha_i}(\tau) \overline{\mathcal{K}_{0,\tilde\alpha_i}(\tau)}  \ .
\end{equation}
where $\mathcal{K}_{\lambda,\alpha}$ are the $\widehat{\mathfrak{u}}(1)_\alpha$ characters \eqref{eq:DefK} and $\alpha_i := \frac12  \bm{o}_i^2$ and $\tilde\alpha_i := -\frac12  \bm{\tilde o}_i^2$, $i=1,\ldots,D$,  are positive integers.

The remaining contributions to the partition function~\eqref{eq:toruspart2} arise from the lattice points $\Gamma_{D,D}$ that do not reside in the sublattice $O_L \oplus O_R$. Let $n+1$ be the index of the sublattice  $O_L \oplus O_R$ embedded in $\Gamma_{D,D}$, and let $\bm{\rho}_1,\ldots,\bm{\rho}_n \in \Gamma_{D,D}$ be $n$ lattice vectors that represent, together with a vector $\bm\rho_0 \in O_L \oplus O_R$, the cosets of the quotient $\Gamma_{D,D} / O_L \oplus O_R$. Then any lattice point $\bm{p}\in\Gamma_{D,D}$, $\bm{p}\notin O_L \oplus O_R$, obeys $\bm p -\bm{\rho}_a \in O_L \oplus O_R$ for some $a \in \{ 1,\ldots n\}$. Furthermore, $\bm{\rho}_a$ has the following expansion in terms of the generators $\bm{o}_i$ and $\bm{\tilde o}_i$
\begin{equation}
  \bm{\rho}_a = \rho_{a,1} \bm{o}_1 + \ldots + \rho_{a,n} \bm{o}_n + \tilde\rho_{a,1} \bm{\tilde o}_1 + \ldots + \tilde\rho_{a,n} \bm{\tilde o}_n \ , \qquad
  a=0,\ldots, n \ ,
\end{equation}
where the coefficients $\rho_{a,i}$ and $\tilde\rho_{a,i}$ are rational numbers. Without loss of generality, we can always assume that these coefficients reside in the interval $[0,1)$, because adding an integer to any of these coefficients does not change the equivalence class of $\bm{\rho}_a$ in the quotient $\Gamma_{D,D} / O_L \oplus O_R$. Moreover, since $\bm{\rho}_a\in\Gamma_{D,D}$, the product of $\bm{\rho}_a$ with generators $\bm{o}_i$ and $\bm{\tilde o}_j$ must be an integer. Thus, using $\bm{o}_i^2 = 2 \alpha_i$ and $\bm{\tilde o}_i^2 = -2 \tilde\alpha_i$, we obtain
\begin{equation}
  \bm{\rho}_a \bm{o}_i = 2 \rho_{a,i} \alpha_i  \in \mathbb{Z} \ , \qquad
  \bm{\tilde\rho}_a \bm{o}_i = - 2 \tilde\rho_{a,i} \tilde\alpha_i  \in \mathbb{Z} \ ,
\end{equation}  
which implies 
\begin{equation}
\begin{aligned}
    \rho_{a,i} &= \frac{\lambda_{a,i}}{2\alpha_i} \quad \text{with} \quad \lambda_{a,i} \in \{ 0,\ldots, 2\alpha_i-1\} \ , \\
    \tilde\rho_{a,i} &= \frac{\tilde\lambda_{a,i}}{2\tilde\alpha_i} \quad \text{with} \quad \tilde\lambda_{a,i} \in \{ 0,\ldots,2\tilde\alpha_i-1 \} \ .
\end{aligned}    
\end{equation}
As such, the $D$-dimensional vector $(\rho_{a,1},\cdots,\rho_{a,D})$ in the basis of $\bm{o}_i$ is an element of $O_L^*$, and likewise $(\tilde\rho_{a,1},\cdots,\tilde\rho_{a,D})\in O_R^*$, where $O_L^*$ and $O_R^*$ are the dual lattices.

We now have all the ingredients at hand to rewrite the lattice sum in the partition function~\eqref{eq:toruspart2}. In addition to the contribution from the sublattice~$O_L \oplus O_R$ \eqref{eq:Zsublattice}, the contribution from remaining lattice points in $\Gamma_{D,D}$ is obtained by summing over $\bm{\rho}_a + (O_L \oplus O_R)$ for all $a=1,\ldots,n$. All in all, we obtain 
\begin{equation} \label{eq:ZTNrat2}
\begin{aligned}
   Z_{T^D}^\text{rat}(\tau) &= Z^{O_L \oplus O_R}_{T^D}(\tau) 
   + \frac1{|\eta(\tau)|^{2D}} \sum_{a=1}^n  \prod_{i=1}^D
   \left( \sum_{k\in\mathbb{Z}} q^{\alpha_i \left(k + \frac{\lambda_{a,i}}{2\alpha_i}\right)^2} \right)\!\!\!
   \left( \sum_{k\in\mathbb{Z}} {\bar q}^{\tilde \alpha_i \left(k+  \frac{\tilde\lambda_{a,i}}{2\tilde\alpha_i}\right)^2} \right) \\
   &=  \sum_{a=0}^n \prod_{i=1}^D \mathcal{K}_{\lambda_{a,i},\alpha_i}(\tau) \,
   \overline{\mathcal{K}_{\tilde\lambda_{a,i},\tilde\alpha_i}(\tau)} \ .
\end{aligned}   
\end{equation}
This is indeed of the general form~\eqref{eq:ZTNrat}.

To summarize, in order to arrive at the expansion~\eqref{eq:ZTNrat2}, we need to first determine the generators $\bm{o}_i$ and $\bm{\tilde o}_i$, $i=1,\ldots,D$, of the  sublattice $O_L\oplus O_R$. In the second step we determine the lattice vectors $\bm{\rho}_a$, $a=1,\ldots,n$. This data readily determines the expansion~\eqref{eq:ZTNrat2}. Note that the expression~\eqref{eq:ZTNrat2} is not unique because it depends on the choice of the sublattice $O_L \oplus O_R$, according to eq.~\eqref{eq:OrthSub}. Different choices for $O_L \oplus O_R$ give rise to different expansion~\eqref{eq:ZTNrat} into distinct products of the characters~\eqref{eq:DefK} of extensions of $\widehat{\mathfrak{u}}(1)$ current algebra. 

\subsection{Comparison of extended chiral algebras of toroidal CFTs}\label{sec:RTCFTcomp}
As discussed in the previous subsection, a criterium for a toroidal CFT to be rational is the existence of even $D$-dimensional lattices $\Gamma_{L,0}$ and $\Gamma_{0,R}$ that are mutually orthogonal such that the orthogonal sum $\Gamma_{L,0}\oplus\Gamma_{0,R}$ is a sublattice of the even self-dual lattice $\Gamma_{D,D}$. Conversely, starting from the lattices $\Gamma_{L,0}$ and $\Gamma_{0,R}$, there is a procedure to reconstruct the embedding self-dual lattice~$\Gamma_{D,D}$. This process is called gluing, and it is a well-known lattice construction \cite{nikMR544937,MR525944}.

In fact, the construction of $\Gamma_{D,D}$ may also be done through gluing primitive sublattices $\Gamma_{L,0}$ and $\Gamma_{0,R}=(\Gamma_{L,0})^\perp\cap\Gamma_{D,D}$.
On the level of the partition function this amounts to
\begin{equation} \label{eq:ZNblocks}
   Z_{T^D}^\text{rat}(\tau) = \frac{1}{|\eta(\tau)|^{2D}} \sum_{\bm{v_L}+\bm{v_R} \in \Gamma_G}\sum_{\substack{\bm{w_L} \in \Gamma_{L,0} \\ \bm{w_R} \in \Gamma_{0,R}} } q^{\frac{1}{2}(\bm{w_L}+\bm{v_L})^2} \bar{q}^{\frac{1}{2}(\bm{w_R}+\bm{v_{R}})^2} \ ,
\end{equation}
where the first sum is over a finite set $\Gamma_G \subset \Gamma_{D,D}$  of glue vectors $\bm{v_L}+\bm{v_R}$ that are decomposed into the vectors $\bm{v_L} \in \Gamma_{L,0}^*$ and $\bm{v_R} \in \Gamma_{0,R}^*$.

We define the holomorphic and anti-holomorphic lattice theta functions for $\Gamma_{L,0}$ and $\Gamma_{0,R}$ as
\begin{equation} \label{eq:thetalattice}
  \Theta_{\bm v}^{\Gamma_{L,0}}(\tau) 
  = \frac{1}{\eta(\tau)^D}\sum_{\bm{w_L} \in \Gamma_{L,0}} q^{\frac{1}{2} (\bm{w_L}+{\bm v})^2} \ , \quad
  \overline{\Theta}_{\bm{v}}^{\Gamma_{0,R}}(\bar\tau)
  = \frac{1}{\eta(\bar{\tau})^D}\sum_{\bm{w_R} \in \Gamma_{0,R}} \bar{q}^{\frac{1}{2} (\bm{w_R}+{\bm v})^2} \ ,
\end{equation}
The partition function~\eqref{eq:ZNblocks} then takes the form
\begin{equation} \label{eq:ZNblocks2}
    Z_{T^D}^\text{rat}(\tau) = \sum_{\bm{v_L}+\bm{v_R} \in \Gamma_G}  
    \Theta_{\bm{v_L}}^{\Gamma_{L,0}}(\tau) \,
    \overline{\Theta}_{\bm{v_R}}^{\Gamma_{0,R}}(\bar\tau) \ .
\end{equation}
The lattice theta functions $\Theta_{\bm{v_L}}^{\Gamma_{L,0}}$ and $\overline{\Theta}_{\bm{v_R}}^{\Gamma_{0,R}}$ are the characters of the extended chiral and anti-chiral algebra of the rational toroidal CFT, respectively. 
The lattices $\Gamma_{L,0}$ and $\Gamma_{0,R}$ are the root lattices with respect to the extended chiral and anti-chiral algebras. The chiral algebra associated to the lattice $\Gamma_{L,0}$ is a $W$-algebra  $\mathcal{W}(2, |\bm{\kappa}_1|^2, \dots, |\bm{\kappa}_D|^2)$ labelled in terms of the squares of the generators $\sqrt{2} \bm{\kappa}_i$, $i=1,\ldots,D$, of the lattice~$\Gamma_{L,0}$, see for instance ref.~\cite{Blumenhagen:2009zz}. Analogously, the characters of the anti-chiral $W$-algebra $\mathcal{W}(2, |\bm{\tilde\kappa}_1|^2, \dots, |\bm{\tilde\kappa}_D|^2)$ are associated to the generators $\sqrt{2}\bm{\tilde\kappa}_i$, $i=1,\ldots,D$, of the $\Gamma_{0, R}$ lattice.

If the sublattice $O_L$ is isometric to $\Gamma_{L,0}$, then the lattice generators $\sqrt{2}\bm{\kappa}_i$ can be identified with the generators $\bm{o}_i$ --- see above eq. \eqref{eq:OrthSub}. In this case, the holomorphic characters of the extended chiral algebra defined in eq.~\eqref{eq:thetalattice} factorize into the characters~\eqref{eq:DefK} of extensions of the $\widehat{\mathfrak{u}}(1)$ current algebra, namely
\begin{equation} \label{eq:factchar}
  \Theta_{\bm{w_L}}^{O_L}(\tau) = \prod_{i=1}^D \mathcal{K}_{\lambda_i,|\bm{\kappa}_i^2|}(\tau)\ ,
\end{equation}
where the labels $\lambda_i$ are obtained from the expansion $\bm{w_L} = \frac1{\bm{\kappa}_i^2}\left( \lambda_1 \bm{o}_i + \ldots + \lambda_D \bm{o}_D \right)$. If in addition the sublattice $O_R$ is isometric to $\Gamma_{0,R}$, then our construction coincides with the gluing construction \cite{nikMR544937,MR525944}.

On the other hand, if the lattice $O_L$ is not isometric to the lattice $\Gamma_{L,0}$, then the holomorphic characters $\Theta_{\bm{w_L}}^{\Gamma_{L,0}}(\tau)$ do not decompose further into products of characters $\mathcal{K}_{\lambda,\alpha}(\tau)$ of extended current algebras $\widehat{\mathfrak{u}(1)}_\alpha$, but instead become a sum of such products. In this case the more general construction described in Section~\ref{sec:RTCFTpar} must be applied in order to arrive at the decomposition~\eqref{eq:ZTNrat} for the holomorhic part of the partition function. Similar considerations apply to the anti-holomorhic sector.

From a geometric perspective, factorizations of the form~\eqref{eq:factchar} of the extended chiral characters occur if the metric $\bm G$ of the target space torus $T^D$ is both rational and diagonal, and if the $B$-field $\bm B$ vanishes. The torus $T^D$ is then a product $S^1 \times \ldots \times S^1$ of $D$ circles as a Riemannian manifold, and $\mathcal{K}_{\lambda_i,|\bm{\kappa}_i^2|}(\tau)$ are the characters of the extended current algebra $\widehat{\mathfrak{u}(1)}_{|\bm{\kappa}_i^2|}$ associated to each rational circle $S^1$. However, factorizations~\eqref{eq:factchar} can also accur for more general target space tori and choices of the $B$-field.

\section{The Decomposition Construction} \label{sec:T2Par}
In this section we present an explicit construction to calculate the decomposition~\eqref{eq:ZTNrat2} of the partition function of rational toroidal CFTs. After reviewing rational toroidal CFTs with target space $T^2$ and their partition functions, we explicitly derive our construction for rational toroidal CFTs with two-dimensional target space tori. However, the described procedure works for any dimensions, and  we show that the decomposition construction for two-dimensional target space tori --- without any further modification --- carries over to rational toroidal CFTs with target space tori of arbitrary dimension.

\subsection{Rational conformal field theories with target space $T^2$}\label{sec3geom}
A toroidal CFT with target space $T^2$ is unambiguously determined by its complex structure modulus $u$ and complexified K\"ahler modulus $t$, which both take values in the Siegel upper half space $\mathcal{H} = \left\{ z \in \mathbb{C} \,\middle|\, \operatorname{Im} z >0 \right\}$, i.e.,
\begin{equation} \label{eq:utH}
  u,t \in \mathcal{H} \ .
\end{equation}  
In terms of these moduli the metric $\bm{G}$ and the skew-symmetric $B$-field $\bm{B}$ of the target space torus $T^2$ read
\begin{equation} \label{eq:GBT2}
  \bm{G}= \frac{2 t_2}{u_2} \begin{pmatrix}1& u_1\\u_1& |u|^2 \end{pmatrix} \ ,  \quad
  \bm{B}=  2 \begin{pmatrix}0& t_1\\ -t_1& 0 \end{pmatrix} \ ,
\end{equation}
where $u_1 =\operatorname{Re} u$,  $u_2 =\operatorname{Im} u$, $t_1 =\operatorname{Re} t$,  and $t_2 =\operatorname{Im} t$. Furthermore, the squares of the right- and left-moving momenta $\bm{p_L}$ and $\bm{p_R}$ are given by (e.g., ref.~\cite{Blumenhagen:2013fgp})
\begin{equation} \label{eq:plpr}	
	\bm{p_L}^2= \frac{1}{2t_2 u_2} |m_2-u m_1+ \bar{t}(n_1+u n_2)|^2 \ , \quad
	\bm{p_R}^2= \frac{1}{2t_2 u_2} |m_2-u m_1+ t(n_1+u n_2)|^2 \ .
\end{equation}
Here the lattice quantum numbers $\bm{m} = (m_1, m_2)\in\mathbb{Z}^2$ and $\bm{n}=(n_1,n_2)\in\mathbb{Z}^2$ entering $\bm{p_L}^2$ and $\bm{p_R}^2$ determine the partition function $Z_{T^2}(\tau; u, t)$, which with eq.~\eqref{eq:toruspart} becomes a function of $u$ and $t$.

The toroidal CFT with target space $T^2$ is rational, if the metric $\bm{G}$ and the $B$-field $\bm{B}$ are rational, c.f., eq.~\eqref{eq:GBrat}. From eq.~\eqref{eq:GBT2} it is straight forward to see that this implies that both complex structure modulus $u$ and the complexified K\"ahler modulus $t$ take values in a quadratic algebraic number field $\mathbb{Q}(\sqrt{-D})$ for some positive square-free integer $D$ \cite{Moore:1989vd,Wendland:2000ye,Hosono:2002yb}. That is to say, altogether with eq.~\eqref{eq:utH} we have
\begin{equation} \label{eq:Ratut}
\begin{aligned}
    &u = a + b \sqrt{-D} \ ,  &&t = c + d \sqrt{-D} \ , \\
    &a,c \in \mathbb{Q}, \quad b,d \in \mathbb{Q}_{>0}, \quad &&D \in \mathbb{Z}_{>0}\ \text{and $D$ square-free} \ ,
\end{aligned}    
\end{equation}
which geometrically implies that both the complex structure modulus $u$ and the complexified K\"ahler modulus $\rho$ describe a torus of complex multiplication.

Inserting the expression \eqref{eq:Ratut} into eq.~\eqref{eq:plpr}, we find
\begin{equation}
\begin{aligned}
   \bm{p_L}^2&= \frac{\left(-a m_1+m_2 + c n_1 + (ac +Dbd)n_2\right)^2}{2 bd D} + \frac{\left(b m_1 + d n_1 + (ad- bc  ) n_2\right)^2}{2 bd}  \ , \\
   \bm{p_R}^2&=\frac{\left(- a m_1+m_2 + c n_1 + (ac -Dbd)n_2\right)^2}{2 bd D} + \frac{\left(-b m_1 + d n_1 + (b c + ad) n_2\right)^2}{2 bd}  \ ,
\end{aligned}	
\end{equation}
which --- by multiplying the numerators and denominators of the summands of these rational expressions with the smallest common denominators --- can be rewritten into
\begin{equation} \label{eq:pLpR_Rat}
\begin{aligned}
   \bm{p_L}^2&= \frac{\left(\bm{a}_1^T \bm{m} + \bm{b}_1^T \bm{n}\right)^2}{2a_1} + \frac{\left(\bm{a}_2^T \bm{m} + \bm{b}_2^T \bm{n}\right)^2}{2a_2}  \ ,
   && a_i \in \mathbb{Z}_{>0} \ , \ \bm{a}_i,\bm{b}_i \in \mathbb{Z}^2 \ , \\
   \bm{p_R}^2&= \frac{\left(\bm{\tilde a}_1^T \bm{m} + \bm{\tilde b}_1^T \bm{n}\right)^2}{2\tilde a_1} 
      + \frac{\left(\bm{\tilde a}_2^T \bm{m} + \bm{\tilde b}_2^T \bm{n}\right)^2}{2\tilde a_2}  \ , 
        \quad &&  \tilde a_i \in  \mathbb{Z}_{>0} \ , \ \bm{\tilde a}_i,\bm{\tilde b}_i\in \mathbb{Z}^2 \ .
\end{aligned}
\end{equation}
The positive integral constants $a_i,  \tilde a_i$, $i=1,2$, and the integral constant vectors $\bm{a}_i, \bm{b_i}, \bm{\tilde a}_i, \bm{\tilde b_i}$, $i=1,2$, are functions of the rational numbers $a, b, c, d$, which define the moduli $u$ and $t$ in eq.~\eqref{eq:Ratut}. Inserting the above equations into the general expression~\eqref{eq:toruspart}, we explicitly see that the partition function $Z_{T^2}^\text{rat}(\tau;u,t)$ is of the form~\eqref{eq:toruspart2}. Eq.~\eqref{eq:pLpR_Rat} holds for target spaces $T^d$ in any dimension.

\subsection{Minimally extended $\widehat{\mathfrak{u}}(1)$ current algebra decomposition}
Our starting point to work out the decomposition of the toroidal partition with target space $T^2$ are the squares of the left- and right-moving momenta as stated in eq.~\eqref{eq:pLpR_Rat}. We collect the defining vectors and denominators in these equations in the $4\times 4$-matrices $\bm{H}$ and $\bm{D}$
\begin{equation} \label{eq:DefMats}
  \bm H = \begin{pmatrix} 
    \bm{a}_1^T & \bm{b}_1^T \\   \bm{a}_2^T & \bm{b}_2^T \\ 
    \bm{\tilde a}_1^T & \bm{\tilde b}_1^T \\   \bm{\tilde a}_2^T & \bm{\tilde b}_2^T 
   \end{pmatrix} \ , \qquad
   \bm D = \begin{pmatrix}
     \phantom{-}2a_1 \\ & \phantom{-}2a_2 \\ && -2\tilde a_1 \\ &&& -2\tilde a_2  
    \end{pmatrix} \ .
\end{equation}       
By construction both matrices have only integral entries and their determinants are non-vanishing.

As the integral intersection pairing $\bm{\Sigma}: \mathbb{Z}^4 \times \mathbb{Z}^4 \to \mathbb{Z}$ of the even self-dual lattice~$\Gamma_{2,2}$ is induced by the norm $\norm{(\bm{m},\bm{n})}^2 = \bm{p_L}^2(\bm m,\bm n) - \bm{p_R}^2(\bm m,\bm n)$ for $(\bm{m},\bm{n})\in\mathbb{Z}^4$ in terms of the momenta squares~\eqref{eq:pLpR_Rat}, we find that the pairing $\bm\Sigma$ is explicitly given by the integral symmetric $4\times 4$-matrix
\begin{equation} \label{eq:DefSigma}
  \bm\Sigma = \bm{H}^T \bm{D}^{-1} \bm{H} \in \operatorname{Mat}_4(\mathbb{Z}) \ .
\end{equation}
The partition function~\eqref{eq:toruspart2} for rational toroidal CFT with target space $T^2$ reads
\begin{equation} \label{eq:ZratT2}
    Z^\text{rat}_{T^{2}}(\tau;u,t) = \frac{1}{|\eta(\tau)|^{2}} \!\!\sum_{\bm{m},\bm{n}\in \mathbb{Z}^2} 
    e^{  \begin{pmatrix} \bm{m}, \bm{n} \end{pmatrix} \bm H^T \bm T(\tau) \bm{H} \begin{pmatrix} \bm{m}\\ \bm{n} \end{pmatrix} } \ .
\end{equation}
with the diagonal matrix
\begin{equation}
 \bm T(\tau) = 
    \begin{pmatrix} 
       \frac{i \pi \tau}{2a_1} \\ &  \frac{i \pi \tau}{2a_2} \\&&  -\frac{i \pi \bar\tau}{2\tilde a_1} \\ &&& -\frac{i \pi \bar\tau}{2\tilde a_2}
     \end{pmatrix} \ .
\end{equation}   

Since the lattice $\Gamma_{2,2}$ is unimodular, the inverse matrix $\bm{\Sigma}^{-1}$ is integral as well. Thus, from eq.~\eqref{eq:DefSigma} we get
\begin{equation}
  \bm\Sigma^{-1} = \bm{H}^{-1} \bm{D} \left(\bm{H}^{-1}\right)^T  \in \operatorname{Mat}_4(\mathbb{Z}) \ ,
\end{equation}
which implies
\begin{equation} \label{eq:Drel}
  \bm{D} = \bm{H} \left( \bm\Sigma^{-1} \bm{H}^{T} \right) \ .
\end{equation}
Due to the integral entries of the matrices $\bm{H}$ and $\bm{\Sigma}^{-1}$, we observe that the columns of the diagonal matrix $\bm{D}$ are integral linear combinations of the columns of the matrix~$\bm{H}$ in terms of the columns of the integral matrix $\left( \bm\Sigma^{-1} \bm{H}^{T} \right)$. 

We can now interpret the columns of the matrix $\bm{H}$ as generators of the lattice~$\Gamma_{2,2}$ together with the intersection paring given by the inverse matrix $\bm{D}^{-1}$. Furthermore, as a consequence of the integral linear relation~\eqref{eq:Drel} the columns of the matrix $\bm{D}$ realize a sublattice $O_L \oplus O_R$ of $\Gamma_{2,2}$ because it has mutually orthogonal generators. This sublattice $O_L \oplus O_R$ also respects the chain of inclusions~\eqref{eq:OrthSub} that are required for the decomposition of the partition function into the characters $\mathcal{K}_{\lambda,\alpha}$ of the form~\eqref{eq:ZTNrat2}. Evaluating the sum in the partition function~\eqref{eq:ZratT2} only on the sublattice $O_L \oplus O_R$ yields the contribution~\eqref{eq:Zsublattice}, which reads
\begin{equation} \label{eq:PartSubD}
\begin{aligned}
  Z^{O_L \oplus O_R}_{T^2}(\tau;u,t) &= 
  \frac{1}{|\eta(\tau)|^{2}} \!\!\sum_{\bm{m},\bm{n}\in \mathbb{Z}^2} 
    e^{  \begin{pmatrix} \bm{m}, \bm{n} \end{pmatrix} \bm D \bm T(\tau) \bm{D} \begin{pmatrix} \bm{m}\\ \bm{n} \end{pmatrix} } 
    = \prod_{i=1}^2 \mathcal{K}_{0,a_i}(\tau) \overline{\mathcal{K}_{0,\tilde a_i}(\tau)}  \ .
\end{aligned}    
\end{equation}

We now construct the remaining terms of the partition function. Let the index of the sublattice generated by the columns of $\bm{D}$ within the lattice generated by the columns of $\bm{H}$ be $n+1$. Then --- as discussed in Section~\ref{sec:RTCFTpar} --- we need to construct $n+1$ vectors $\bm\rho_0, \ldots, \bm\rho_n$ representing the $n+1$ distinct cosets of the quotient of the lattice of $\bm{H}$ by the sublattice of $\bm{D}$. Note that the index of this sublattice is given by
\begin{equation} \label{eq:IndexD}
  n+1 = \frac{\det \bm{D}}{\det \bm{H}}=2^4\frac{a_1 a_2 \tilde a_1 \tilde a_2}{\det \bm{H}} = {\det \bm{H}}  \ .
\end{equation}
Here the last equality is a consequence of $\det\bm{\Sigma} =1$ together with eq.~\eqref{eq:DefSigma}. 

Without loss of generality let us assume that the integral matrix $\bm{H}$ is in (column) Hermite normal form\footnote{If the matrix $\bm{H}$ is not in this form, it can always be brought into Hermite normal form because for any integral matrix $\bm{M}$ there exists a uni-modular matrix $\bm{U}$ such that the matrix product $\bm{M}\bm{U} = \bm{H}$ is in Hermite normal form.  Moreover, since the multiplication with a uni-modular matrix realizes a lattice automorphism, any lattice generated by the column vectors of $\bm{M}$ can be represented by a matrix $\bm{H}$ in Hermite normal form. To transform a matrices $\bm{M}$ into Hermite normal form $\bm{H}$, there are standard algorithms, which are implemented in most modern computer algebra software packages.}
\begin{equation} \label{eq:Hermite}
  \bm{H}  = 
  \begin{pmatrix} 
    h_{11} & 0 & 0 &0 \\ h_{21} & h_{22} & 0 & 0 \\ h_{31} & h_{32} & h_{33} & 0 \\ h_{41} & h_{42} & h_{43} & h_{44} 
  \end{pmatrix} \quad \in \ \operatorname{Mat}_4(\mathbb{Z}) \ ,
\end{equation}  
and the diagonal entries of the inverse matrix $\bm{H}^{-1}$ read
\begin{equation}
  \bm{H}^{-1} =
   \begin{pmatrix} 
    \frac1{h_{11}} & \cdots & \cdots & \cdots \\ 0 & \frac1{h_{22}} &  \cdots & \cdots \\ 0 & 0 & \frac1{h_{33}} & \cdots \\ 0 & 0 & 0 & \frac1{h_{44}} 
  \end{pmatrix} \ .
\end{equation}
Due to eq.~\eqref{eq:Drel} the matrix product $\bm{H}^{-1} \bm{D}$ is integral with the diagonal entries
\begin{equation}
    \bm{H}^{-1} \bm{D} =  
   \begin{pmatrix} 
    \frac{2a_1}{h_{11}} & \cdots & \cdots & \cdots \\ 0 & \frac{2a_2}{h_{22}} &  \cdots & \cdots \\ 0 & 0 & -\frac{2\tilde a_1}{h_{33}} & \cdots \\ 
    0 & 0 & 0 & -\frac{2\tilde a_2}{h_{44}} 
  \end{pmatrix} \quad \in \ \operatorname{Mat}_4(\mathbb{Z}) \ ,
 \end{equation}   
which in particular establishes the important result 
\begin{equation} \label{eq:defss}
   s_1=\frac{2a_1}{h_{11}}, \ s_2=\frac{2a_2}{h_{22}}, \ \tilde s_1=\frac{2\tilde a_1}{h_{33}}, \  \tilde s_2=\frac{2\tilde a_2}{h_{44}}  \in \mathbb{Z} \ .
\end{equation}   

Next we determine the lattice vectors $\bm{\rho}_0, \bm{\rho}_1, \ldots, \bm{\rho}_n$ representing the cosets associated to the sublattice of the diagonal matrix $\bm{D}$. Based on the Hermite normal form~\eqref{eq:Hermite} of the matrix $\bm{H}$, we construct the set of lattice vectors
\begin{equation} \label{eq:DefC}
   \mathcal{C} = \left\{ \bm{H} \begin{pmatrix} r_1 \\ r_2 \\ \tilde r_1 \\ \tilde r_2 \end{pmatrix}\,
   \middle| \,  r_i \in \{ 0, \ldots, s_i-1\},\ \tilde r_i \in \{ 0, \ldots, \tilde s_i-1\} \ \ \text{for} \ \ i=1,2 
    \right\} \ ,
\end{equation}   

Now we proof that all elements of the set $\mathcal{C}$ mutually represent distinct cosets. To show this, let us pick two lattice vectors $\bm{\rho}, \bm{\sigma} \in \mathcal{C}$, which are obtained from the integers $r_1,r_1' \in \{ 0, \ldots, s_i-1\}, \ldots, \tilde{r}_1,\tilde{r}_1' \in \{ 0, \ldots, s_i-1\}$, respectively. These two lattice vectors represent the same coset if and only if their difference $\bm{\rho} - \bm{\sigma}$ is a point on the sublattice generated by the columns of the matrix~$\bm{D}$. This difference explicitly reads
\begin{equation}
  \bm{\rho}-\bm{\sigma} = \bm{H} \begin{pmatrix} \Delta r_1  \\ \Delta r_2  \\ \Delta \tilde r_1  \\ \Delta \tilde r_2  \end{pmatrix} 
  = 
  \begin{pmatrix}
     h_{11} \Delta r_1 \hfill \\
     h_{21} \Delta r_1 + h_{22} \Delta r_2 \hfill \\
     h_{31} \Delta r_1 + h_{32} \Delta r_2 + h_{33} \Delta r_3 \hfill \\
     h_{41} \Delta r_1 + h_{42} \Delta r_2 + h_{43} \Delta r_3 + h_{44} \Delta r_4 \hfill
  \end{pmatrix}  \ ,
\end{equation}
where $\Delta r_1 = r_1 - r_1', \ldots, \Delta\tilde r_2 = \tilde r_2 - \tilde r_2'$. Note that --- due to the relations~\eqref{eq:defss} --- $|\Delta r_1| < s_1 = \frac{2a_1}{h_{11}}$, and therefore we have that the absolute value of the first entry of the vector $\bm{\rho}-\bm{\sigma}$ is smaller than $2a_1$. Hence, looking at the difference entry, the vector $\bm{\rho} -\bm{\sigma}$ can only be on the sublattice of $\bm{D}$, i.e., can only be an integral multiple of the first column vector of $\bm{D}$, if $\Delta r_1 = 0$. If $\Delta r_1 = 0$ then we look at the second entry of the vector $\bm{\rho} -\bm{\sigma}$. For $\Delta r_1=0$ the term proportional to $h_{21}$ vanishes, and we can repeat the same argument as for the first entry. Namely, in this case the second entry is smaller in absolute value than $a_2$. Thus, by considering the first and the second entry, we find that  $\bm{\rho} -\bm{\sigma}$ can only be on the sublattice if both $\Delta r_1 = 0$ and $\Delta r_2 = 0$. Repeating this argument again inductively for the remaining entries of $\bm{\rho} -\bm{\sigma}$, we conclude that $\bm{\rho} - \bm{\sigma}$ is only a point on the sublattice, if and only if $\Delta r_1 = \Delta r_2 = \Delta \tilde r_1 = \Delta \tilde r_2 = 0$, which is equivalent to $\bm{\rho} = \bm{\sigma}$. Thus, altogether we conclude that all the lattice vector in the set $\mathcal{C}$ defined in eq.~\eqref{eq:DefC}, describe distinct cosets with respect to the sublattice of $\bm{D}$. 

According to eqs.~\eqref{eq:IndexD} and \eqref{eq:defss}, the cardinality of the set $\mathcal{C}$ is given by
\begin{equation}
   \left| \strut \mathcal{C}\right| = s_1 s_2 \tilde s_1 \tilde s_2 = 2^4\frac{a_1 a_2 \tilde a_1 \tilde a_2}{h_{11} h_{22} h_{33} h_{44}} = n +1 \ ,
\end{equation}  
which shows that the vectors in the set $\mathcal{C}$ represent all the cosets with respect to the sublattice $\bm{D}$. We denote the $n+1$ elements of $\mathcal{C}$ in the following by the lattice vectors $\bm{\rho}_0, \ldots, \bm{\rho}_{n}$ (where $\bm{\rho}_0$ is the zero vector of the set $\mathcal{C}$).

With the vectors $\bm{\rho}_a$, $a=0,\ldots, n$, at hand, we can now express the decomposition of the entire partition function $Z_{T^2}^\text{rat}(\tau; u, t)$ explicitly in terms of the characters $\mathcal{K}_{\lambda, \alpha}$ as 
\begin{equation} \label{eq:DecompFinal}
Z^\text{rat}_{T^2}(\tau;u,t) 
    = \sum_{a=0}^n \prod_{i=1}^2 \mathcal{K}_{\rho_{a,i},a_i}(\tau) \overline{\mathcal{K}_{\rho_{a,i+2},\tilde a_i}(\tau)}  \ ,
\end{equation}
where $\rho_{a,i}$ are the entries of the column vectors $\bm{\rho}_a$, $a=0,\ldots, n$. Note that the contribution~\eqref{eq:PartSubD} of the sublattice corresponds to the trivial coset given by the null vector $\bm{\rho}_0$. 

Let us stress that the presented construction for the decomposition~\eqref{eq:DecompFinal} of rational toroidal CFTs with target space $T^2$ applies without any further modification to any rational toroidal CFTs with target space $T^D$ for any dimension $D$. Namely, starting with the partition function~$Z_{T^D}^\text{rat}(\tau)$ in the form~\eqref{eq:toruspart2}, we can read off a $2D\times 2D$-dimensional matrix $\bm{H}$ and the $2D\times 2D$-dimensional diagonal matrix $\bm{D} =\operatorname{Diag}\left( 2a_1, \ldots, 2a_D, -2\tilde a_1, \ldots, -2\tilde a_D \right)$. As illustrated for the target space $T^2$, the columns of $\bm{D}$ also generate a sublattice of the lattice generated by the columns of $\bm{H}$ for general target space dimension $D$. By calculating the Hermite normal form of $\bm{H}$, we read off (analogously as for $D=2$) the character expansion of the form~\eqref{eq:ZTNrat2}. We illustrate this construction explicitly with an example of a rational toroidal CFT with target space $T^3$ in the following section.

\section{Examples} \label{sec:Examples}
In this section we illustrate the general decomposition construction described in the previous section with concrete representative examples. We focus on rational toroidal CFTs with two- and three-dimensional target space tori.

\subsection{Target space $T^2$: Extended $\widehat{\mathfrak{su}}(3)_{1}$ chiral algebra}
Consider the $\widehat{\mathfrak{su}}(3)_{1}$ diagonal partition function with the affine Lie algebra $\widehat{\mathfrak{su}}(3)_{1}$ at level one as its extended chiral algebra. The lattices $\Gamma_{L,0}$ and $\Gamma_{0,R}$ are both the root lattice of $\mathfrak{su}(3)$. This can be realized by setting both the complex structure and the complexified K\"ahler modulus of target space torus $T^2$ to $u=t= -\frac{1}{2} + \frac{\sqrt{-3}}{2}$. The associated partition function explicitly reads
\begin{multline}\label{su3partitionfunction}
	Z_{T^2}^{\widehat{\mathfrak{su}}(3)_{1}}(\tau) = \frac{1}{|\eta(\tau)|^4} \sum_{\bm{m},\bm{n}  \in \mathbb{Z}^2} q^{\frac{1}{12} (m_1+2m_2 - n_1 + 2n_2)^2} q^{\frac{1}{4} (m_1+n_1)^2} \\
	\cdot \bar{q}^{\frac{1}{12} (m_1+2m_2-n_1-n_2)^2} \bar{q}^{\frac{1}{4} (m_1-n_1+n_2)^2} \ .
\end{multline}
From eq.~\eqref{eq:toruspart2} we read off the constants
\begin{equation}
\begin{aligned}
   a_1&=\tilde a_1=3 \ , \ & a_2 &=\tilde a_2 = 1 \ , \ &
   \bm{a}_1&= \bm{\tilde a}_1 = \begin{pmatrix} 1 \\ 2 \end{pmatrix} \ , \ &
   \bm{a}_2&= \bm{\tilde a}_2 = \begin{pmatrix} 1 \\ 0 \end{pmatrix} \ ,  \\
   \bm{b}_1&=  \begin{pmatrix} -1 \\ 2 \end{pmatrix} \ , &
   \bm{b}_2&=  \begin{pmatrix} 1 \\ 0 \end{pmatrix} \ , & 
   \bm{\tilde b}_1&=  \begin{pmatrix} -1 \\ -1 \end{pmatrix} \ , &
   \bm{\tilde b}_2&=  \begin{pmatrix} -1 \\ 1 \end{pmatrix} \ , 
\end{aligned}   
\end{equation}    
which determine the diagonal $4\times 4$-matrix $\bm{D}$ and an integral $4\times 4$-matrix according to eq.~\eqref{eq:DefMats}. The latter matrix has the Hermite normal form
\begin{equation}
	\bm{H}= \begin{pmatrix}
		1&0&0&0\\
		1&2&0&0\\
		1&0&3&0\\
		1&0&1&2
	\end{pmatrix}\ .
\end{equation}
Note that the integers $a_1, a_2, \tilde a_1, \tilde a_2$ fulfill the relation~\eqref{eq:IndexD}, and with eq.~\eqref{eq:defss} give rise to the integers
\begin{equation}
  s_1 = 6 \ , \quad s_2 = 1 \ , \quad \tilde s_1 = 2 \ , \quad \tilde s_2 = 1 \ ,
\end{equation}   
such that the partition function decomposes into
\begin{equation}
	Z_{T^2}^{\widehat{\mathfrak{su}}(3)_{1}}(\tau) = \sum_{r_1 = 0}^5 \sum_{\tilde r_1=0}^1 
	 \mathcal{K}_{r_1,3}(\tau)\, \mathcal{K}_{r_1,1}(\tau) 
	 \overline{\mathcal{K}_{r_1 + 3\tilde r_1,3}(\tau)} \, \overline{\mathcal{K}_{r_1 + \tilde r_1,1}(\tau)} \ .
\end{equation}
Using the symmetries~\eqref{eq:Ksym} of the characters $\mathcal{K}_{\lambda,\alpha}$ --- which for instance imply $\mathcal{K}_{2,1} \equiv \mathcal{K}_{0,1}$, $\mathcal{K}_{5,3} \equiv \mathcal{K}_{1,3}$, $\mathcal{K}_{4,3} \equiv \mathcal{K}_{2,3}$, we arrive at the explicit expansion
\begin{equation}\label{su3thetaexpansion}
\begin{aligned}	
	Z_{T^2}^{\widehat{\mathfrak{su}}(3)_{1}}(\tau) = &\phantom{+}\mathcal{K}_{0,1} \mathcal{K}_{0,3} \bar{\mathcal{K}}_{0,1} \bar{\mathcal{K}}_{0,3} +\mathcal{K}_{1,1} \mathcal{K}_{3,3} \bar{\mathcal{K}}_{0,1} \bar{\mathcal{K}}_{0,3}+2\mathcal{K}_{1,1} \mathcal{K}_{1,3} \bar{\mathcal{K}}_{1,1} \bar{\mathcal{K}}_{1,3}\\ 
	&+2\mathcal{K}_{0,1} \mathcal{K}_{2,3} \bar{\mathcal{K}}_{1,1} \bar{\mathcal{K}}_{1,3}+2\mathcal{K}_{1,1} \mathcal{K}_{1,3} \bar{\mathcal{K}}_{0,1} \bar{\mathcal{K}}_{2,3}+2\mathcal{K}_{0,1} \mathcal{K}_{2,3} \bar{\mathcal{K}}_{0,1} \bar{\mathcal{K}}_{2,3}\\ 
	&+\mathcal{K}_{0,1} \mathcal{K}_{0,3} \bar{\mathcal{K}}_{1,1} \bar{\mathcal{K}}_{3,3} +\mathcal{K}_{1,1} \mathcal{K}_{3,3} \bar{\mathcal{K}}_{1,1} \bar{\mathcal{K}}_{3,3}\\ 
	=&\phantom{+} |\chi_{100}(\tau)|^2 + |\chi_{010}(\tau)|^2 + |\chi_{001}(\tau)|^2  \ ,
\end{aligned}
\end{equation}
with 
\begin{equation}\label{eq:su3latticethetafunction}
\begin{aligned}	
	\chi_{100}(\tau)&=\mathcal{K}_{0,1}(\tau) \mathcal{K}_{0,3}(\tau) + \mathcal{K}_{1,1}(\tau) \mathcal{K}_{3,3}(\tau) \ ,\\
	\chi_{010}(\tau)&=\mathcal{K}_{1,1}(\tau) \mathcal{K}_{1,3}(\tau) + \mathcal{K}_{0,1}(\tau) \mathcal{K}_{2,3}(\tau) \ ,\\
	\chi_{001}(\tau)&=\mathcal{K}_{1,1}(\tau) \mathcal{K}_{1,3}(\tau) + \mathcal{K}_{0,1}(\tau) \mathcal{K}_{2,3}(\tau) \ .
\end{aligned}
\end{equation}
Here $\chi_{100}(\tau)$, $ \chi_{010}(\tau)$, $\chi_{001}(\tau)$ are the irreducible specialized characters of the affine Lie algebra~$\widehat{\mathfrak{su}}(3)_{1}$ at level one. These characters are obtained from the Kac--Weyl character for $\widehat{\mathfrak{su}}(3)_{1}$ upon taking a certain limit as for instance introduced in ref.~\cite{DiFrancesco:1997nk}.

The Gram matrices $\bm{G_L}$ and $\bm{G_R}$ for the chiral and anti-chiral lattices $\Gamma_{L,0}$ and $\Gamma_{0,R}$ are calculated to be the Cartan matrix of the simple Lie algebra~$\mathfrak{su}(3)$, i.e., 
\begin{equation} \label{intersectionsu3}
\bm{G_L} = \bm{G_R} =\begin{pmatrix}
	\phantom{-}2&-1\\
	-1& \phantom{-}2
\end{pmatrix} \ .
\end{equation}
Therefore, generators of the lattices $\Gamma_{L,0}$ and $\Gamma_{0,R}$ cannot be mutually orthogonal, which implies that the specialized characters of $\widehat{\mathfrak{su}}(3)_{1}$ do not factor into characters of the extended current algebras $\widehat{u}(1)_\alpha$. Instead they can only be a sum of such characters, as given in eq.~\eqref{eq:su3latticethetafunction}. These particular expansions arise from the maximally diagonal sublattices $O_L$ and $O_R$ of $\Gamma_{L,0}$ and $\Gamma_{0,R}$ with mutually orthogonal generators with the Gram matrices
\begin{equation}
\bm{G}_{O_L}=\bm{G}_{O_R} =\begin{pmatrix}
	2&0\\
	0&6
\end{pmatrix} \ .
\end{equation}

Note that the decompositions~\eqref{eq:su3latticethetafunction} can easily be calculated. For instance, the vacuum character $\chi_{100}(\tau)$ enjoys the expansion
\begin{equation}
   \chi_{100}(\tau)= \frac{1}{\eta(\tau)^2} \sum_{m_1, m_2 \in \mathbb{Z}} q^{\frac{1}{4} m_1^2} q^{\frac34 (m_1 + 2m_2)^2} \ ,
\end{equation}
which by substituting the summation index $m_1$ by $2 m_1' + \alpha$ for $\alpha \in \{ 0,1 \}$ and by replacing $m_2$ by $m_1' + m_2'$ yields
\begin{equation}
	\chi_{100}(\tau)= \frac{1}{\eta(\tau)^2} \!\!\sum_{\substack{m_1', m_2' \in \mathbb{Z}\\\alpha \in \set{0, 1}}} 
	\!\! q^{\left(m_1'+ \frac{\alpha}{2}\right)^2} q^{3\left(m_2' + \frac{3 \alpha}{6}\right)^3}\\
	= \mathcal{K}_{0,1}(\tau) \mathcal{K}_{0,3}(\tau) + \mathcal{K}_{1,1}(\tau) \mathcal{K}_{3,3}(\tau) \ .
\end{equation}
This is in agreement with eq.~\eqref{eq:su3latticethetafunction}.

\subsection{Target space $T^2$: Moduli $u=t= \frac{1}{2}+\frac{\sqrt{-5}}{2}$}
In this example we consider the target space torus $T^2$ with complex structure modulus and complexified K\"ahler modulus both set to $u=t= \frac{1}{2}+\frac{\sqrt{-5}}{2}$. This example is generic in the sense that the maximally extended chiral algebra is not given by an affine semi-simple Lie algebra, but instead by a more general W-algebra. The partition function of this rational toroidal CFT is determined by the data
\begin{equation}
\begin{aligned}
   a_1&=\tilde a_1=5 \ , \ & a_2 &=\tilde a_2 = 1 \ , \ &
   \bm{a}_1&= \bm{\tilde a}_1 = \begin{pmatrix} 1 \\ -2 \end{pmatrix} \ , \ &
   \bm{a}_2&= \bm{\tilde a}_2 = \begin{pmatrix} 1 \\ 0 \end{pmatrix} \ ,  \\
   \bm{b}_1&=  \begin{pmatrix} -1 \\ -3 \end{pmatrix} \ , &
   \bm{b}_2&=  \begin{pmatrix} 1 \\ 0 \end{pmatrix} \ , & 
   \bm{\tilde b}_1&=  \begin{pmatrix} -1 \\2 \end{pmatrix} \ , &
   \bm{\tilde b}_2&=  \begin{pmatrix} -1 \\ -1 \end{pmatrix} \ ,
\end{aligned}   
\end{equation}    
which gives rise to the $4\times 4$-matrix $\bm{H}$ in Hermite normal form
\begin{equation}
\bm{H}=
\begin{pmatrix}
1&0&0&0\\
0&1&0&0\\
6&5&10&0\\
1&0&0&2 
\end{pmatrix} \ .
\end{equation}
Applying the described decomposition construction we find
\begin{equation}
\begin{aligned}
	Z_{T^2}^\text{rat}(\tau;u,t)=
	&\phantom{+}\mathcal{K}_{0, 1} \mathcal{K}_{0, 5} \bar{\mathcal{K}}_{0, 1} \bar{\mathcal{K}}_{0, 5} + 
	\mathcal{K}_{0, 1} \mathcal{K}_{5, 5} \bar{\mathcal{K}}_{0, 5} \bar{\mathcal{K}}_{1, 1} + 
	2 \,\mathcal{K}_{1, 1} \mathcal{K}_{4, 5} \bar{\mathcal{K}}_{0, 1} \bar{\mathcal{K}}_{1, 5} \\  
	&+ 2\, \mathcal{K}_{1, 1} \mathcal{K}_{1, 5} \bar{\mathcal{K}}_{1, 1} \bar{\mathcal{K}}_{1, 5}
	+  2\, \mathcal{K}_{0, 1} \mathcal{K}_{2, 5} \bar{\mathcal{K}}_{0, 1} \bar{\mathcal{K}}_{2, 5} + 
	2\, \mathcal{K}_{0, 1} \mathcal{K}_{3, 5} \bar{\mathcal{K}}_{1, 1} \bar{\mathcal{K}}_{2, 5} \\
	&+ 2\, \mathcal{K}_{1, 1} \mathcal{K}_{2, 5} \bar{\mathcal{K}}_{0, 1} \bar{\mathcal{K}}_{3, 5} 
	+2\, \mathcal{K}_{1, 1} \mathcal{K}_{3, 5} \bar{\mathcal{K}}_{1, 1} \bar{\mathcal{K}}_{3, 5} 
	+ 2\, \mathcal{K}_{0, 1} \mathcal{K}_{4, 5} \bar{\mathcal{K}}_{0, 1} \bar{\mathcal{K}}_{4, 5} \\
	&+ 2\, \mathcal{K}_{0, 1} \mathcal{K}_{1, 5} \bar{\mathcal{K}}_{1, 1} \bar{\mathcal{K}}_{4, 5} + 
	\mathcal{K}_{1,1} \mathcal{K}_{0, 5} \bar{\mathcal{K}}_{0, 1} \bar{\mathcal{K}}_{5, 5} + 
	\mathcal{K}_{1, 1} \mathcal{K}_{5, 5} \bar{\mathcal{K}}_{1, 1} \bar{\mathcal{K}}_{5, 5} \ .
\end{aligned}
\end{equation}

\subsection{Target space $T^2$:  Moduli $u=t= \frac{1}{4} + \frac{\sqrt{-3}}{4}$}
In this example we consider the target space torus $T^2$ with complex structure modulus and complexified K\"ahler modulus both set to $u=t= \frac{1}{4} + \frac{\sqrt{-3}}{4}$. This example illustrates that generically the summands in the decompositions factor into characters $\mathcal{K}_{\lambda,\alpha}$ with distinct indices $\alpha$. For these moduli the partition function of the rational toroidal CFT is given by
\begin{equation}
\begin{aligned}
   a_1&=3 \ , \ & a_2 & = 1 \ , \ & \tilde{a}_1&=12 \ , \ & \tilde{a}_2 & = 4  \  , & \\
     \bm{a}_1&=  \begin{pmatrix} 1 \\ -4 \end{pmatrix} \ , &
   \bm{\tilde{a}}_1&=  \begin{pmatrix} 2 \\ -8 \end{pmatrix} \ , & 
   \bm{a}_2&=  \begin{pmatrix} 1 \\ 0 \end{pmatrix} \ , &
   \bm{\tilde a}_2&=  \begin{pmatrix} 2 \\ 0 \end{pmatrix} \ , 
   \\
   \bm{b}_1&=  \begin{pmatrix} -1 \\ -1 \end{pmatrix} \ , &
   \bm{\tilde b}_1&=  \begin{pmatrix} -2 \\ 1 \end{pmatrix} \ , & 
   \bm{ b}_2&=  \begin{pmatrix} 1 \\ 0 \end{pmatrix} \ , &
   \bm{\tilde b}_2&=  \begin{pmatrix} -2 \\ -1 \end{pmatrix} \ .
\end{aligned}   
\end{equation}   
The associated $4\times 4$-matrix $\bm{H}$  in Hermite normal form reads
\begin{equation}
\bm{H}=
\begin{pmatrix}
1&0&0&0\\
0&1&0&0\\
5&3&6&0\\
3&1&2&8
\end{pmatrix} \ ,
\end{equation}
which determines the decomposition
\begin{equation}
\begin{aligned}
  Z_{T^2}^\text{rat}(\tau;u,t)=
  &\phantom{+}\mathcal{K}_{0, 1} \mathcal{K}_{0, 3} \bar{\mathcal{K}}_{0, 4} \bar{\mathcal{K}}_{0, 12} 
  +2\,\mathcal{K}_{0, 1} \mathcal{K}_{1, 3} \bar{\mathcal{K}}_{1, 4} \bar{\mathcal{K}}_{1, 12} 
  +2\,\mathcal{K}_{1, 1} \mathcal{K}_{1, 3} \bar{\mathcal{K}}_{2, 4} \bar{\mathcal{K}}_{2, 12} \\ 
  &+2\,\mathcal{K}_{0, 1} \mathcal{K}_{2, 3} \bar{\mathcal{K}}_{2, 4} \bar{\mathcal{K}}_{2, 12}
  +2\,\mathcal{K}_{1, 1} \mathcal{K}_{2, 3} \bar{\mathcal{K}}_{3, 4} \bar{\mathcal{K}}_{1, 12}
  +2\,\mathcal{K}_{1, 1} \mathcal{K}_{0, 3} \bar{\mathcal{K}}_{1, 4} \bar{\mathcal{K}}_{3, 12} \\
  &+2\,\mathcal{K}_{0, 1} \mathcal{K}_{3, 3} \bar{\mathcal{K}}_{3, 4} \bar{\mathcal{K}}_{3, 12} 
  +\mathcal{K}_{1, 1} \mathcal{K}_{3, 3} \bar{\mathcal{K}}_{4, 4} \bar{\mathcal{K}}_{0, 12}
  +2\,\mathcal{K}_{1, 1} \mathcal{K}_{1, 3} \bar{\mathcal{K}}_{0, 4} \bar{\mathcal{K}}_{4, 12}\\
  &+2\,\mathcal{K}_{0, 1} \mathcal{K}_{2, 3} \bar{\mathcal{K}}_{4, 4} \bar{\mathcal{K}}_{4, 12}
  +2\,\mathcal{K}_{1, 1} \mathcal{K}_{2, 3} \bar{\mathcal{K}}_{1, 4} \bar{\mathcal{K}}_{5, 12} 
  +2\, \mathcal{K}_{0, 1} \mathcal{K}_{1, 3} \bar{\mathcal{K}}_{3, 4} \bar{\mathcal{K}}_{5, 12} \\
  &+2\,\mathcal{K}_{0, 1} \mathcal{K}_{0, 3} \bar{\mathcal{K}}_{2, 4} \bar{\mathcal{K}}_{6, 12} 
  +2\,\mathcal{K}_{1, 1} \mathcal{K}_{3, 3} \bar{\mathcal{K}}_{2, 4} \bar{\mathcal{K}}_{6, 12}
  +2\,\mathcal{K}_{0, 1} \mathcal{K}_{1, 3} \bar{\mathcal{K}}_{1, 4} \bar{\mathcal{K}}_{7, 12}\\
  &+2\,\mathcal{K}_{1, 1} \mathcal{K}_{2, 3} \bar{\mathcal{K}}_{3, 4} \bar{\mathcal{K}}_{7, 12} 
  +2\,\mathcal{K}_{0, 1} \mathcal{K}_{2, 3} \bar{\mathcal{K}}_{0, 4} \bar{\mathcal{K}}_{8, 12}
  +2\,\mathcal{K}_{1, 1} \mathcal{K}_{1, 3} \bar{\mathcal{K}}_{4, 4} \bar{\mathcal{K}}_{8, 12}\\
  &+2\,\mathcal{K}_{0, 1} \mathcal{K}_{3, 3} \bar{\mathcal{K}}_{1, 4} \bar{\mathcal{K}}_{9, 12} 
  +2\,\mathcal{K}_{1, 1} \mathcal{K}_{0, 3} \bar{\mathcal{K}}_{3, 4} \bar{\mathcal{K}}_{9, 12}
  +2\,\mathcal{K}_{1, 1} \mathcal{K}_{1, 3} \bar{\mathcal{K}}_{2, 4} \bar{\mathcal{K}}_{10, 12}\\
  &+2\,\mathcal{K}_{0, 1} \mathcal{K}_{2, 3} \bar{\mathcal{K}}_{2, 4} \bar{\mathcal{K}}_{10, 12}
  +2\,\mathcal{K}_{1, 1} \mathcal{K}_{2, 3} \bar{\mathcal{K}}_{1, 4} \bar{\mathcal{K}}_{11, 12}
  +2\,\mathcal{K}_{0, 1} \mathcal{K}_{1, 3} \bar{\mathcal{K}}_{3, 4} \bar{\mathcal{K}}_{11, 12}\\
  &+\mathcal{K}_{1, 1} \mathcal{K}_{3, 3} \bar{\mathcal{K}}_{0, 4} \bar{\mathcal{K}}_{12, 12}
  +\mathcal{K}_{0, 1} \mathcal{K}_{0, 3} \bar{\mathcal{K}}_{4, 4} \bar{\mathcal{K}}_{12, 12} \ .
\end{aligned}
\end{equation}

Note that the modular $\operatorname{PSL}(2,\mathbb{Z})$-transformation $z \mapsto \frac{z-1}z$ maps both moduli $u=t= \frac{1}{4} + \frac{\sqrt{-3}}{4}$ to $u = t =\sqrt{-3}$, which realizes the target space $T^2$ as a product of circles $S^1\times S^1$ with radii $\sqrt{2}$ and $\sqrt{6}$ and vanishing $B$-field. Thus, due to this duality relation the discussed partition function also enjoys the expansion
\begin{multline}
  Z_{T^2}^\text{rat}(\tau;u,t)=Z_{T^2}^\text{rat}(\tau;\sqrt{-3},\sqrt{-3}) \\
  =
  \left( \mathcal{K}_{0, 1} \bar{\mathcal{K}}_{0, 1} +\mathcal{K}_{1, 1} \bar{\mathcal{K}}_{1, 1} \right) 
  \left( \mathcal{K}_{0, 3} \bar{\mathcal{K}}_{0, 3} +2\mathcal{K}_{1, 3} \bar{\mathcal{K}}_{1, 3} 
  +2\mathcal{K}_{2, 3} \bar{\mathcal{K}}_{2, 3} +\mathcal{K}_{3, 3} \bar{\mathcal{K}}_{3, 3}  \right) \ ,
\end{multline}
where the individual factors are the decompositions~\eqref{eq:ZS1ratDec} of the rational circular CFTs with radii $\sqrt{2}$ and $\sqrt{6}$, respectively.

\subsection{Target space $T^2$:  Moduli $u= \frac{1}{2} + \sqrt{-1}$ and $t=\sqrt{-1}$}
In this example we consider the target space torus $T^2$ and $B$-field for which the complex structure modulus $u = \frac12 +\sqrt{-1}$ and the complexified K\"ahler modulus $t =\sqrt{-1}$ are distinct. These moduli yield a rational toroidal CFT with a non-factorizable target space torus $T^2$. Such CFTs possess a non-generic $\mathbb{Z}_2$~symmetry, as for instance recently studied in ref.~\cite{Forste:2024zjt}. For these moduli the CFT is determined by
\begin{equation}
\begin{aligned}
   a_1&=\tilde a_1=4 \ , \ & a_2 &=\tilde a_2 = 4 \ , \ &
   \bm{a}_1&= \bm{\tilde a}_1 = \begin{pmatrix} 1 \\ -2 \end{pmatrix} \ , \ &
   \bm{a}_2&= \bm{\tilde a}_2 = \begin{pmatrix} 2 \\ 0 \end{pmatrix} \ ,  \\
   \bm{b}_1&=  \begin{pmatrix} 0 \\ -2 \end{pmatrix} \ , &
   \bm{b}_2&=  \begin{pmatrix} 2 \\ 1 \end{pmatrix} \ , & 
   \bm{\tilde b}_1&=  \begin{pmatrix} 0 \\ 2 \end{pmatrix} \ , &
   \bm{\tilde b}_2&=  \begin{pmatrix} -2 \\ -1 \end{pmatrix} \ ,
\end{aligned}   
\end{equation}  
which gives rise to the $4\times 4$-matrix $\bm{H}$ in Hermite normal form
\begin{equation}
\bm{H}=
\begin{pmatrix}
1&0&0&0\\
0&1&0&0\\
1&4&8&0\\
4&7&0&8
\end{pmatrix} \ .
\end{equation}
From this data we extract that the integers~\eqref{eq:defss} become
\begin{equation}
  s_1 = s_2 = 8 \ ,  \quad \tilde s_1 = \tilde s_2 = 1 \ ,
\end{equation}
such that there are $64$ cosets~\eqref{eq:DefC}. The resulting decomposition reads
\begin{equation}
\begin{aligned}
	Z_{T^2}^\text{rat}(\tau;u,t)
	=&\ \mathcal{K}_{0, 4} \mathcal{K}_{0, 4} \bar{\mathcal{K}}_{0, 4} \bar{\mathcal{K}}_{0, 4} 
	+4\,\mathcal{K}_{1, 4} \mathcal{K}_{4, 4} \bar{\mathcal{K}}_{0, 4} \bar{\mathcal{K}}_{1, 4}
	+4\, \mathcal{K}_{3, 4} \mathcal{K}_{3, 4} \bar{\mathcal{K}}_{1, 4} \bar{\mathcal{K}}_{1, 4}\\
	&+4\, \mathcal{K}_{0, 4} \mathcal{K}_{2, 4} \bar{\mathcal{K}}_{0, 4} \bar{\mathcal{K}}_{2, 4}
	+8\, \mathcal{K}_{1, 4} \mathcal{K}_{2, 4} \bar{\mathcal{K}}_{1, 4} \bar{\mathcal{K}}_{2, 4}
	+4\,\mathcal{K}_{2, 4} \mathcal{K}_{2, 4} \bar{\mathcal{K}}_{2, 4} \bar{\mathcal{K}}_{2, 4} \\
	&+4\,\mathcal{K}_{3, 4} \mathcal{K}_{4, 4} \bar{\mathcal{K}}_{0, 4} \bar{\mathcal{K}}_{3, 4}
	+8\,\mathcal{K}_{1, 4} \mathcal{K}_{3, 4} \bar{\mathcal{K}}_{1, 4} \bar{\mathcal{K}}_{3, 4}
	+8\,\mathcal{K}_{2, 4} \mathcal{K}_{3, 4} \bar{\mathcal{K}}_{2, 4} \bar{\mathcal{K}}_{3, 4} \\
	&+4\,\mathcal{K}_{1, 4} \mathcal{K}_{1, 4} \bar{\mathcal{K}}_{3, 4} \bar{\mathcal{K}}_{3, 4}
	+2\,\mathcal{K}_{0,4} \mathcal{K}_{4, 4} \bar{\mathcal{K}}_{0, 4} \bar{\mathcal{K}}_{4, 4} 
	+4\,\mathcal{K}_{0, 4} \mathcal{K}_{1, 4} \bar{\mathcal{K}}_{1, 4} \bar{\mathcal{K}}_{4, 4}\\
	&+4\,\mathcal{K}_{2, 4} \mathcal{K}_{4, 4} \bar{\mathcal{K}}_{2, 4} \bar{\mathcal{K}}_{4, 4}
	+4\,\mathcal{K}_{0, 4} \mathcal{K}_{3, 4} \bar{\mathcal{K}}_{3, 4} \bar{\mathcal{K}}_{4, 4}
	+\mathcal{K}_{4, 4} \mathcal{K}_{4, 4} \bar{\mathcal{K}}_{4, 4} \bar{\mathcal{K}}_{4, 4} \ .
\end{aligned}
\end{equation}
\subsection{Target space $T^3$: Extended $\widehat{\mathfrak{su}}(4)_{1}$ chiral algebra}
The rational toroidal CFT with the $\widehat{\mathfrak{su}}(4)_{1}$ diagonal partition function arises from a three-dimensional target space torus $T^3$, for which the three-dimensional lattices $\Gamma_{L,0}$ and $\Gamma_{0,R}$ are the root lattices of the simple Lie algebra $\mathfrak{su}(4)$. This CFT is associated to the target space metric $\bm{G}$ and the $B$-field~$\bm{B}$ with the rational entries
\begin{equation}
  \bm{G}=\begin{pmatrix}
  \phantom{+}1&-\frac{1}{2}&\phantom{+}0\\
  -\frac{1}{2}&\phantom{+}1&-\frac{1}{2}\\
  \phantom{+}0&-\frac{1}{2}&\phantom{+}1
  \end{pmatrix}\ , \qquad
  {\bm{B}}=\begin{pmatrix}
  \phantom{+}0&\phantom{+}\frac{3}{2}&\phantom{+}1\\ 
  -\frac{3}{2}&\phantom{+}0&\phantom{+}\frac{1}{2}\\
  -1&-\frac{1}{2}&\phantom{+}0
  \end{pmatrix} \ ,
\end{equation}
from which --- following ref.~\cite{Wendland:2000ye}  --- we determine the defining data
\begin{equation}
\begin{aligned}
   a_1&=\tilde a_1 =1 \ , \ & a_2 &= \tilde a_2  = 3 \ , \ & a_3&=\tilde a_3=6 \ , \\
   \bm{a}_1&=  \begin{pmatrix} 1 \\ 0 \\0 \end{pmatrix} \ , &
   \bm{a}_2&=  \begin{pmatrix} 1 \\ 2 \\ 0 \end{pmatrix} \ , &
   \bm{a}_3&=  \begin{pmatrix} 1 \\ 2 \\ 3 \end{pmatrix} \ , \\
   \bm{b}_1&=  \begin{pmatrix} \phantom{+}1 \\ -2 \\ -1 \end{pmatrix} \ , &
   \bm{b}_2&=  \begin{pmatrix} \phantom{+}3 \\ \phantom{+}0 \\ -3 \end{pmatrix} \ , &
   \bm{b}_3&=  \begin{pmatrix} 6 \\ 0 \\ 0 \end{pmatrix} \ , \\
   \bm{\tilde{a}}_1&=  \begin{pmatrix} 1 \\ 0 \\ 0 \end{pmatrix} \ , & 
   \bm{\tilde a}_2&=  \begin{pmatrix} 1 \\ 2 \\ 0 \end{pmatrix} \ , & 
   \bm{\tilde a}_3&=  \begin{pmatrix} 1 \\ 2 \\ 3 \end{pmatrix} \ ,\\
   \bm{\tilde b}_1&=  \begin{pmatrix} -1 \\ -1 \\ -1 \end{pmatrix} \ , & 
   \bm{\tilde b}_2&=  \begin{pmatrix} \phantom{+}3 \\ -3 \\ -1 \end{pmatrix} \ , &
   \bm{\tilde b}_3&=  \begin{pmatrix} \phantom{+}6 \\ \phantom{+}0 \\ -4 \end{pmatrix}\ .
\end{aligned}   
\end{equation}   
The resulting integral $6\times 6$-matrix $\bm{H}$ in Hermite normal form becomes
\begin{equation}
\bm{H}=\begin{pmatrix}
1&0&0&0&0&0\\
1&2&0&0&0&0\\
1&2&3&0&0&0\\
0&0&0&1&0&0\\
0&0&0&1&2&0\\
9&6&3&4&8&12
\end{pmatrix}.
\end{equation}
The integers $a_1, a_2, a_3,  \tilde a_1, \tilde a_2, \tilde a_3$ fulfill the relation~\eqref{eq:IndexD}, and with eq.~\eqref{eq:defss} give rise to the integers
\begin{equation}
  s_1 = 2 \ , \quad s_2 = 3 \ , \quad s_3=4 \ ,   \quad \tilde s_1 = 2 \ , \quad \tilde s_2 = 3 \ , \quad \tilde s_3=1
\end{equation}   
such that the partition function decomposes into
\begin{multline}
	Z_{T^3}^{\widehat{\mathfrak{su}}(4)_{1}}(\tau) = \sum_{r_1,\tilde r_1 = 0}^1 \sum_{ r_2,\tilde r_2=0}^2  \sum_{ r_3=0}^3
	 \mathcal{K}_{r_1,1}(\tau)\, \mathcal{K}_{r_1 + 2 r_2,3}(\tau) \, \mathcal{K}_{r_1 + 2 r_2 +3 r_3, 6}(\tau) \\
	 \cdot \overline{\mathcal{K}_{\tilde r_1,1}(\tau)} \, \overline{\mathcal{K}_{\tilde r_1 + 2 \tilde r_2,3}(\tau)} \,  \overline{\mathcal{K}_{9 r_1 +6 r_2 + 3 r_3 + 4 \tilde r_1 + 8\tilde r_2, 6}(\tau)}\ ,
\end{multline}
which can be written in the specialized characters of the affine Lie algebra $\widehat{\mathfrak{su}}(4)_{1}$ as
\begin{equation}
 Z_{T^3, \, \widehat{\mathfrak{su}}(4)_{1}}(\tau)
   =|\chi_{1000}(\tau)|^2 + |\chi_{0100}(\tau)|^2 + |\chi_{0010}(\tau)|^2 +|\chi_{0001}(\tau)|^2 
\end{equation}
with
\begin{equation} \label{su4character}
\begin{aligned}
  \chi_{1000}(\tau)&=\mathcal{K}_{0,1} \mathcal{K}_{0,3} \mathcal{K}_{0,6} 
      + \mathcal{K}_{1,1} \mathcal{K}_{3,3} \mathcal{K}_{0,6} + 2 \mathcal{K}_{1,1} \mathcal{K}_{1,3} \mathcal{K}_{4,6} 
      + 2 \mathcal{K}_{0,1} \mathcal{K}_{2,3} \mathcal{K}_{4,6} \ ,\\   
   \chi_{0100}(\tau)&=\mathcal{K}_{1,1} \mathcal{K}_{1,3} \mathcal{K}_{1,6} 
      + \mathcal{K}_{0,1} \mathcal{K}_{2,3} \mathcal{K}_{1,6} + \mathcal{K}_{0,1} \mathcal{K}_{0,3} \mathcal{K}_{3,6} \\
      & \qquad\qquad\qquad + \mathcal{K}_{1,1} \mathcal{K}_{3,3} \mathcal{K}_{3,6} 
      + \mathcal{K}_{1,1} \mathcal{K}_{1,3} \mathcal{K}_{5,6} + \mathcal{K}_{0,1} \mathcal{K}_{2,3} \mathcal{K}_{5,6}\ , \\ 
  \chi_{0010}(\tau)&=\chi_{0100}(\tau) \ , \\ 
  \chi_{0001}(\tau)&= 2  \mathcal{K}_{1,1}  \mathcal{K}_{1,3}  \mathcal{K}_{2,6} 
      + 2  \mathcal{K}_{0,1}  \mathcal{K}_{2,3}  \mathcal{K}_{2,6} + \mathcal{K}_{0,1}  \mathcal{K}_{0,3}  \mathcal{K}_{6,6} 
      + \mathcal{K}_{1,1}  \mathcal{K}_{3,3}  \mathcal{K}_{6,6} \ .
\end{aligned}
\end{equation}
These specialized characters of $\widehat{\mathfrak{su}}(4)_{1}$ arise from the Kac--Weyl character for $\widehat{\mathfrak{su}}(4)_{1}$ upon taking a certain limit as for instance detailed in ref.~\cite{DiFrancesco:1997nk}. As the lattices $\Gamma_{L,0}$ and $\Gamma_{0,R}$ are the root lattices of ${\mathfrak{su}}(4)$, their intersection form becomes the Cartan martix of the simple Lie algebra $\mathfrak{su}(4)$, namely
\begin{equation}
\bm{G_L} = \bm{G_R}=
\begin{pmatrix}
\phantom{+}2&-1&\phantom{+}0\\
-1&\phantom{+}2&-1\\
\phantom{+}0&-1&\phantom{+}2
\end{pmatrix} \ .
\end{equation}
The maximally diagonal sublattices $O_L$ and $O_R$ of $\Gamma_{L,0}$ and $\Gamma_{0,R}$ with mutually orthogonal generators have the $3\times 3$ Gram matrices
\begin{equation}
  \bm{G}_{O_L}=\bm{G}_{O_R}=
\begin{pmatrix}
  2&0&0\\
  0&6&0\\
  0&0&12
\end{pmatrix} \ .
\end{equation}

\section*{Acknowledgements}
We would like to thank 
Alexandros Kanargias
and
Pyry Kuusela
for interesting and useful discussions.
This work is supported by the Cluster of Excellence Precision Physics, Fundamental Interactions, and Structure of Matter (PRISMA+, EXC 2118/1) within the German Excellence Strategy (Project-ID 390831469).

\bigskip
\bibliographystyle{utphys}
\bibliography{JSZ}
\end{document}